\documentstyle[11pt]{article}

\columnwidth\textwidth
\title{On Gauge Invariance and Spontaneous Symmetry Breaking  \thanks{Work
supported by Swiss National Science Foundation}}
\author{A.Aste, G.Scharf\\
Institut f\"ur Theoretische Physik der Universit\"at Z\"urich\\
Winterthurerstrasse 190, CH-8057 Z\"urich, Switzerland
\and M. D\"utsch\thanks{Work supported by
Alexander von Humboldt Foundation}\\
II. Institut f\"ur Theoretische Physik der Universit\"at Hamburg\\
Luruper Chaussee 149, D-22761 Hamburg, Germany}
\date{\today}
\begin{document}
\maketitle
\advance\baselineskip 11 pt

\begin{abstract}
We show how the widely used concept of spontaneous symmetry breaking
can be explained in causal perturbation theory by introducing
a perturbative version of quantum gauge invariance.
Perturbative gauge invariance, formulated exclusively
by means of asymptotic fields, is discussed
for the simple example of Abelian $U(1)$ gauge theory
(Abelian Higgs model). Our findings are
relevant for the electroweak theory, as pointed out elsewhere.
\\

interactions.
\end{abstract}
\newpage
\section{Introduction}
It is quite a common assumption that scalar
QED with massive photons is not a gauge
theory {\em{ in the usual sense}}, because the introduction of a mass
term in the Lagrangean for the
gauge field violates the {\em{classical gauge invariance}} of the theory.
Therefore, a 'Higgs' field with non-vanishing vacuum expectation
value is usually coupled to the photon which then acquires a mass [1].
Proceeding in this way, the local $U(1)$ invariance is not absent,
but 'hidden'.

It is the aim of this paper to demonstrate how massive gauge theories can
be described in the framework of causal perturbation theory [2] by
means of a perturbative version of quantum gauge invariance (25).
Perturbative gauge invariance has the advantage that it provides a
powerful tool for the actual construction of the theory.
We will demonstrate this for the Abelian Higgs model in Sect. 4.
\footnote{We would like to thank Bert Schroer for posing this problem}
Starting from a cubic coupling $\sim A_\mu A^\mu \phi$, gauge invariance
of first order demands the introduction of scalar ghost fields $u, \tilde u$
and of an additional unphysical scalar field $\Phi$ and fixes most of the
cubic couplings. Then, gauge invariance to second order determines the
remaining cubic couplings and requires additional quartic ones. One has to
go to third order to fix the quartic couplings completely.
The resulting couplings contain the Higgs potential which, however, comes
out as a quartic polynomial in the original asymptotic scalar field $\phi$
with vanishing vacuum expectation value $\langle \phi \rangle=0$.
That means, gauge invariance leads us directly to the final theory
'after spontaneous symmetry breaking'. Although we can see the symmetry
breaking in the double-well potential at the end, it plays no direct
role in the construction:
perturbative gauge invariance alone does the job.

The method beautifully works in the more complicated situation of the
electroweak theory, as pointed out in detail elsewhere [9,10].

\section{Gauge Invariance for Massive Gauge Fields}
\subsection{Causal Perturbation Theory}
Our work is best done in the framework of causal perturbation theory, which
has its roots in a classical paper by Epstein and Glaser [2]. In this approach
the $S$-matrix is constructed inductively order by order in the form
\begin{equation}
S(g)=1+\sum_{n=1}^\infty{1\over n!}\int d^4x_1\ldots d^4x_n\,
T_n(x_1,\ldots x_n)g(x_1)\ldots g(x_n),
\end{equation}
where $g(x)$ is a tempered test function that switches the interaction.
The first order (e.g. for QED)
\begin{equation}
T_1 (x) = ie:\bar{\Psi}(x) \gamma^\mu \Psi(x): A_\mu(x)
\end{equation}
must be given in terms of the asymptotic free fields. It is a striking
property of the causal approach that {\em no ultraviolet divergences}
appear, i.e. the $T_n$'s are finite and well defined up to finite
normalization terms. The adiabatic limit $g(x) \rightarrow 1$ has been
shown to exist in purely massive theories in each order [2].

The crucial point in the causal formulation of perturbation theory is that the
usual formal definition of the $T_n$ via simple time-ordering
\begin{equation}
T_n(x_1,...x_n)=T\{T_1(x_1) \cdot ... T_1(x_n)\}
\end{equation}
\begin{equation}
\equiv \sum \limits_\Pi \Theta(x^o_{\Pi_1}-x^o_{\Pi_2}) \cdot ...
\Theta(x^o_{\Pi_{n-1}}-x^o_{\Pi_n})T_1(x_{\Pi_1}) \cdot ... T_1(x_{\Pi_n}),
\end{equation}
where the sum runs over all $n !$ permutations, contains ultraviolet
divergences, therefore there must be an error in the derivation.
Epstein and Glaser proceed more carefully and introduce the following n-point
distributions:
\begin{equation}
A'_n(x_1,\ldots x_n)=\sum_{P_2}\tilde T_{n_1}(X)T_{n-n_1}(Y,x_n),
\end{equation}
\begin{equation}
R'_n(x_1,\ldots x_n)=\sum_{P_2}T_{n-n_1}(Y,x_n)\tilde T_{n_1}(X),
\end{equation}
where the sums run over all partitions
\begin{equation}
P_2:\quad\{x_1,\ldots x_{n-1}\}=X\cup Y,\quad X\ne\emptyset
\end{equation}
into disjoint subsets with $|X|=n_1$, $|Y|\le n-2$. Assuming by induction
that $T_1,...T_{n-1}$ are known, then $A'_n$ and $R'_n$ can be calculated.
One also introduces
\begin{equation}
D_n(x_1,\ldots x_n)=R'_n-A'_n.
\end{equation}
If the sums are extended over all partitions $P_2^0$, including the
empty set $X=\emptyset$, we obtain the distributions
\begin{equation}
A_n(x_1,\ldots x_n)=\sum_{P_2^0}\tilde T_{n_1}(X)T_{n-n_1}(Y,x_n)=
\end{equation}
\begin{equation}
=A'_n+T_n(x_1,\ldots x_n), \label{ad}
\end{equation}
\begin{equation}
R_n(x_1,\ldots x_n)=\sum_{P_2^0}T_{n-n_1}(Y,x_n)\tilde T_{n_1}(X)=
\end{equation}
\begin{equation}
=R'_n+T_n(x_1,\ldots x_n). \label{rd}
\end{equation}
These two distributions are not known by the induction assumption
because they contain the unknown $T_n$. Only the difference
\begin{equation}
D_n=R'_n-A'_n=R_n-A_n \label{dd}
\end{equation}
is known.
We stress the fact that all products of distributions in here are well-defined
because the arguments are disjoint sets of points so that the products are
tensor products of distributions.

One can determine $R_n$ or $A_n$
separately by investigating the support properties of the various
distributions. Causality of the $S$-matrix requires
that $R_n$ is a retarded and $A_n$ an
advanced distribution [2,3]
\begin{equation}
\mbox{supp} \, R_n\subseteq \bar \Gamma^+_{n-1}(x_n),
\quad\mbox{supp} \, A_n\subseteq
\bar \Gamma^-_{n-1}(x_n),
\end{equation}
with
$$
\bar \Gamma^\pm_{n-1}(x)\equiv\{(x_1,\ldots x_{n-1})\>|\>x_j\in\bar V^\pm
(x),\forall j=1,\ldots n-1\},
$$
\begin{equation}
\bar V^\pm(x)=\{y\>|\>(y-x)^2\ge 0,\>\pm (y^0-x^0)\ge 0\}.
\end{equation}
Hence, by splitting of the causal distribution (\ref{dd}) one gets $R_n$
(and $A_n$), and $T_n$ then follows from (\ref{ad}) (or (\ref{rd})). The
$T_n$'s
so obtained are well-defined time-ordered products. Local terms with support
$(x_1=...=x_n)$, originating from a certain ambiguity in the splitting
procedure, might spoil the symmetry of the $T_n$'s in
$x_1,...x_n$, but this minor problem can be removed by subsequent
symmetrization.

To carry out the splitting process, we write (\ref{dd}) in normally ordered
form and split the numerical distributions $d_n^k(x)$, where
$x=(x_1-x_n,...,x_{n-1}-x_n)$
\begin{equation}
D_n(x_1,...x_n)=\sum \limits_{{\cal O}} d_n^{\cal O} (x_1-x_n,...x_{n-1}-x_n)
:{\cal O}(x_1,...x_n):.
\end{equation}
$:{\cal O}:$ is a normally ordered product of external field operators
(Wick monomial).
It is a consequence of translation invariance that $d_n^{\cal O}(x)$
depends only on relative coordinates.

The only nontrivial step in the construction of well-defined time-ordered
products is the splitting of a numerical distribution $d$ with support
in ${\bar \Gamma}^+ \cup {\bar \Gamma}^-$ into a distribution
$r$ with support in ${\bar \Gamma}^+$ and a distribution $a$ with
support in ${\bar \Gamma}^-$. In causal perturbation theory the usual
formal time-ordered products with subsequent renormalization
are replaced by this conceptually simple and
mathematically well-defined procedure. In fact the problem of distribution
splitting was already solved in a general framework by the mathematician
Malgrange in 1960 [4]. Epstein and Glaser used his general result for the
special case of relativistic quantum field theory [2]. A simple solution
for the splitting problem can be found in [3].
\subsection{Gauge Invariance for Massive QED}
Since the above construction of the perturbative $S$-matrix uses only
the asymptotic free fields, we are looking for a formulation of
quantum gauge invariance in terms of these fields.

We discuss first
the simple case of quantum electrodynamics with massive photons.
Let
\begin {equation}
Q \stackrel{def}{=} \int d^3 x (\partial_\mu A^\mu(x) + m \Phi(x)) \stackrel
{\leftrightarrow}{\partial_0} u_a(x)
\end{equation}
be the generator of (free) gauge transformations, called gauge charge for
brevity. $A_\mu$ is the gauge potential in the Feynman gauge,
$u$, $\tilde{u}$ are
fermionic ghost fields and $\Phi$ a neutral scalar, satisfying
the well-known commutation relations
\begin{equation}
[A_\mu^{(\pm)}(x),A_\nu^{(\mp)}(y)]=ig^{\mu \nu} D_m^{(\mp)}(x-y),
\end{equation}
\begin{equation}
\{u^{(\pm)}(x),\tilde{u}^{(\mp)}(y)\}=-iD_m^{(\mp)}(x-y),
\end{equation}
\begin{equation}
[\Phi^{(\pm)}(x),\Phi^{(\mp)}(y)]=-iD_m^{(\mp)}(x-y)
\end{equation}
and all other \{anti-\}commutators vanish. All these fields fulfil the
Klein-Gordon equation with the same mass $m$.
In order to see how the
infinitesimal gauge tranformation acts on the free fields, we calculate
the (anti-)commutators [9]
\begin{equation}
[Q,A_\mu]=i\partial_\mu u \quad , \quad [Q,\Phi]=imu \quad ,
\end{equation}
\begin{equation}
\{Q,u\}=0 \quad , \quad \{Q,\tilde{u}\}= -i \partial_\mu A^\mu-
im \Phi \quad \quad , \quad [Q,\Psi]=0.
\end{equation}
Then we have
\begin{equation}
[Q,T_1(x)] = -e:\bar{\Psi} \gamma^\mu \Psi: \partial_\mu u
\end{equation}
\begin{equation}
=i \partial_\mu (ie:\bar{\Psi} \gamma^\mu \Psi:u) = i \partial_\mu T^\mu_{1/1}
(x).
\end{equation}
Assuming that the operation of commuting with $Q$ commutes with time-ordering,
we obtain
\begin{equation}
[Q,T_n(x_1,...x_n)] = i \sum_{l=1}^{n} \partial_\mu^{x_l} T_{n/l}^{\mu}
(x_1,...x_n) = (sum \, of \, divergences) \quad , \label{dive}
\end{equation}
where $T^\mu_{n/l}$ is a mathematically rigorous version of the time-ordered
product
\begin{equation}
T^\mu_{n/l}(x_1,...,x_n)
\, "=" \, T(T_1(x_1)...T^\mu_{1/1} (x_l)...T_1(x_n)) \quad ,
\end{equation}
constructed by means of the method of Epstein and Glaser described above.
We define (\ref{dive}) to be the condition of gauge invariance [3].
For a fixed $x_l$ we consider from $T_n$ all terms with the external field
operator $A_\mu(x_l)$
\begin{equation}
T_n(x_1,...x_n) = :t^\mu_l(x_1,...x_n) A_\mu(x_l):+...
\end{equation}
(the dots represent terms without $A_\mu(x_l)$).
Then gauge invariance requires
\begin{equation}
\partial_\mu^l [t^\mu_l(x_1,...x_n)u(x_l)]=t^\mu_l(x_1,...x_n) \partial_\mu
u(x_l)
\end{equation}
or
\begin{equation}
\partial_\mu^l t^\mu_l(x_1,...x_n) = 0. {\label{eich}}
\end{equation}
It is an interesting observation that although the photon is massive,
it is not necessary to introduce
a 'Higgs' field to give an explanation for this fact.

\section{Unitarity}
Eq. ({\ref{eich}}) is the usual gauge invariance condition as in
the massless case [3], where no scalar $\Phi$ is needed. Moreover,
$\Phi$ and the ghost fields do not couple at all.
Therefore, we have to explain why the unphysical
fields have been
introduced. The reason is that it allows to prove unitarity of the
$S$-matrix on the physical Hilbert space $H_{phys}$, which is a subspace of the
Fock-Hilbert space $F$ containing also the unphysical ghosts and scalars.

The basic property for unitarity is the nilpotency of the gauge charge
Q
\begin{equation}
Q^2=\frac{1}{2} \{Q,Q\}=0,
\end{equation}
and the Krein structure on the Fock-Hilbert space [5,6,7,8].
Then the physical Fock space can be expressed by the kernel and the
range of Q, namely
\begin{equation}
H_{phys}=ker \, Q \ominus ran \, Q=ker\{Q,Q^+\}.
\end{equation}
This can be most easily seen by realizing the various
field operators on a {\em positive definite} Fock-Hilbert space $F$:
\begin{equation}
A^\mu(x)=(2 \pi)^{-3/2}\sum_{\lambda=0}^{3} \int \frac{d^3k}{\sqrt{2 \omega}}
\Bigl(\epsilon^\mu_\lambda ({\bf{k}}) a_\lambda ({\bf{k}}) e^{-ikx} \pm
(\epsilon^\mu_\lambda ({\bf{k}})
a_\lambda^+({\bf{k}}) e^{+ikx} \Bigr) \quad , \quad
\omega=\sqrt{{\bf{k}}^2+m^2}
\label{vec}
\end{equation}
where $\epsilon_\lambda^\mu$ are four polarization vectors satisfying
\begin{equation}
\epsilon_0^\mu \stackrel{def}{=}  \frac{k^\mu}{m} \quad , \quad
g_{\mu \nu} \epsilon^\mu_\lambda \epsilon^\nu_\kappa = g_{\kappa \lambda} \quad
,
\end{equation}
\begin{equation}
\sum_{\lambda=0}^{3} g_{\lambda \lambda} \epsilon^\mu_\lambda
\epsilon^\nu_\lambda
=g^{\mu \nu} \quad , \quad \epsilon^{\mu*}_\lambda=\epsilon^\mu_\lambda ,
\end{equation}
and we have a minus sign for $\lambda=0$ in (\ref{vec}) to be consistent with
Lorentz invariance.
A similar asymmetry occurs in the ghost sector
\begin{equation}
u(x)=(2 \pi)^{-3/2} \int \frac{d^3k}{\sqrt {2 \omega}} \Bigl( c_2({\bf{k}})
e^{-ikx}+c_1({\bf{k}})^+ e^{ikx} \Bigr),
\end{equation}
\begin{equation}
\tilde{u}(x)=(2 \pi)^{-3/2} \int \frac{d^3k}{\sqrt {2 \omega}}
\Bigl(-c_1({\bf{k}})
e^{-ikx}+c_2({\bf{k}})^+ e^{ikx} \Bigr).
\end{equation}
All creation and annihilation operators satisfy the usual commutation
relations.
Then the proof of unitarity is exactly the same as in [5,6].

We wish to emphasize that we are not forced to represent the gauge potential
in the Feynman gauge as in (\ref{vec}).
If we would not do so, the unphysical particles would acquire
a mass depending on the gauge fixing parameter.
Furthermore, in the case of a massless photon, the above
considerations remain valid with a little exception: The {\em unphysical}
scalar field $\Phi$
would not appear anymore in the gauge charge
$Q$, therefore it would become physical and its mass could
be chosen arbitrarily, or the field could be removed from the theory.

The full power of the above concept shows up if non-abelean gauge fields
are introduced (e.g. in electroweak theory [9,10]). The example
which follows shows some essential features of the more complicated
discussion in case of the elektroweak theory.
For simplicity, we
will demonstrate in the following section
how perturbative gauge invariance fixes all couplings
in the case of an Abelian theory. In a sense, we will {\em{derive}}
the 'Higgs'-potential.

\section{The Abelian Higgs Model}
\subsection{Gauge Invariance at First Order}
Consider the simple case of classical Abelian U(1) gauge theory [11], given
by the Lagrangean
\begin{equation}
{\cal{L}}=(\partial_\mu + ig B_\mu) \varphi^+ (\partial^\mu -ig B^\mu) \varphi
+\mu^{2} \varphi^+ \varphi - \lambda (\varphi^+ \varphi)^2 - \frac{1}{4} F_{\mu
\nu}
F^{\mu \nu},
\end{equation}
\begin{equation}
F^{\mu \nu} = \partial ^ \mu B^\nu - \partial^\nu B^\mu.
\end{equation}
If we assume that the scalar field $\varphi$ develops a vacuum expectation
value $| \langle 0|\varphi|0 \rangle |=v/\sqrt{2}=(\mu^2/2 \lambda)^{1/2}$,
then we obtain in the unitary gauge the Lagrangean
$$
{\cal{L}} = \frac{1}{2} (\partial_\mu \phi)^2 - \frac{1}{2} m_H^2 \phi^2
-\frac{1}{4} (\partial_\mu A_\nu-\partial_\nu A_\mu)^2 +\frac{1}{2}m^2
A_\mu A^\mu \
$$
\begin{equation}
+g^2 v A_\mu A^\mu \phi +\frac{1}{2} g^2 A_\mu A^\mu \phi^2 -
\lambda v \phi^3 - \frac{1}{4} \lambda \phi^4, \label{agree}
\end{equation}
\begin{equation}
m=gv \quad , \quad m_H=\sqrt{2} \mu \quad ,
\end{equation}
where $\phi$ is hermititan and
$(A_\mu,(v+\phi(x))/\sqrt{2})$ are obtained from $(B_\mu,\varphi)$ by a local
$U(1)$ transformation
\begin{equation}
\varphi(x)=\frac{1}{\sqrt{2}} (v+\phi(x))e^{i\xi(x)/v} \quad , \quad
B_\mu(x)=A_\mu(x)+\frac{1}{gv}\partial_\mu\xi(x)
\end{equation}
Now we derive the whole quantum theory in a totally different way.
Our starting point is the first order coupling $\sim A_\mu A^\mu \phi$
of the physical fields $A_\mu$ and $\phi$ with masses $m$ and $m_H$,
respectively.
Furthermore, we introduce the unphysical scalar field $\Phi$
which appears in the gauge charge Q. The latter is still given by (17)
and the guiding principle is the operator gauge invariance (25).
Then a general ansatz for the first order coupling, containing only
trilinear terms in the free fields and leading to a renormalizable theory, is
$$
T_1(x)=igm :[A_\mu A^\mu \phi + \alpha A_\mu A^\mu \Phi + \beta_1
u \tilde{u} \phi + \beta_2 u \tilde{u} \Phi + \beta_3 A_\mu u \partial^\mu
\tilde{u}
$$
\begin{equation}
+\gamma A_\mu (\phi \partial^\mu \Phi-\Phi \partial^\mu \phi)
+\delta_1 \Phi^3 + \delta_2 \Phi^2 \phi + \delta_3 \Phi \phi^2
+\delta_4 \phi^3]:
\end{equation}
We calculate $d_Q T_1=[Q,T_1]$ and obtain
$$
d_Q T_1= -gm:[2 \partial_\mu(u(A^\mu \phi + \alpha A^\mu \Phi))
+\gamma \partial_\mu (u(\phi \partial^\mu \Phi - \Phi \partial^\mu \phi))
$$
$$
+\gamma m \partial_\mu (uA^\mu \phi)
-2 u \partial_\mu A^\mu \phi - 2 u A^\mu \partial_\mu \phi -2 \alpha u
\partial_\mu A^\mu \Phi - 2 \alpha u A^\mu \partial_\mu \Phi
$$
$$
+\alpha m u A_\mu A^\mu +\beta_1 u \partial_\mu A^\mu \phi + \beta_1 m u \Phi
\phi
+\beta_2 u \partial_\mu A^\mu \Phi +\beta_2 m u \Phi^2
$$
$$
\beta_3(\partial^\mu u u \partial_\mu \tilde{u} + A^\mu u \partial_\mu
(\partial_\nu A^\nu + m \Phi ))
$$
$$
+\gamma m^2 u \phi \Phi - \gamma m_H^2 u \phi \Phi - \gamma m u \partial_\mu
A^\mu \phi - 2 \gamma m u A_\mu \partial^\mu \phi
$$
\begin{equation}
+3 \delta_1 m u \Phi^2 +2\delta_2 m u \Phi \phi + \delta_3 mu \phi^2]:
\end{equation}
where we have taken out the derivatives of the ghost fields. Since
$d_Q T_1$ has to be a pure divergence, the terms which are not of this form
must cancel. This fixes most of the free parameters. We immediately obtain
\begin{equation}
T_1=igm:[A^\mu A_\mu \phi + u \tilde{u} \phi -\frac{1}{m}A_\mu
(\phi \partial^\mu \Phi - \Phi \partial^\mu \phi) - \frac{m_H^2}{2 m^2}
\phi \Phi^2 + \delta_4 \phi^3]:
\end{equation}
and
\begin{equation}
d_Q T_1 = -gm:\partial_\mu [(uA^\mu \phi) -\frac{1}{m}
u(\phi \partial^\mu \Phi - \Phi \partial^\mu \phi)]: \stackrel{def}{=}
i \partial_\mu T_{1/1}^\mu. \label{te11}
\end{equation}
Obviously, the quadrilinear couplings in (\ref{agree}) are still missing,
and $\delta_4$ is not yet fixed. Therefore, we have to discuss
gauge invariance at second and third order.

\subsection{Gauge Invariance at Second and Third Order}
Following the inductive construction of Epstein and Glaser, we have to
calculate first the causal distribution
\begin{equation}
D_2(x,y)=T_1(x)T_1(y)-T_1(y)T_1(x)=-A'_2(x,y)+R'_2(x,y).
\end{equation}
The main problem is whether gauge invariance can be preserved in the
distribution splitting. Obviously, $D_2$ is gauge invariant:
$$
d_Q D_2(x,y)= [d_Q T_1(x),T_1(y)] +[T_1(x),d_QT_1(y)]=
$$
\begin{equation}
i\partial^x_\mu [T_{1/1}^\mu (x), T_1(y)]+i\partial^y_\mu [T_1(x),
T_{1/1}^\mu (y)] \stackrel{def}{=} i\partial_\mu^x D_{2/1}^\mu (x,y)
+i\partial_\mu^y D_{2/2}^\mu (x,y).
\end{equation}
Since the retarded part $R_2$ agrees with $D_2$ on the forward light cone
$V^+\setminus \{x=y \}$ and similarly for $R_{2/1}^\mu$,$R_{2/2}^\mu$, gauge
invariance
of $R_2$ can only be violated by local terms $\sim D^a \delta(x-y)$.
But such local terms are precisely the freedom of normalization in the
distribution splitting. If the normalization terms $N_2$,$N_{2/1}^\mu$,
$N_{2/2}^\mu$ can be chosen in such a way that
\begin{equation}
d_Q (R_2 + N_2) = i\partial^x_\mu (R_{2/1}^\mu + N_{2/1}^\mu) +
i\partial_\mu^y (R_{2/2}^\mu + N_{2/2}^{\mu}) \label{gau}
\end{equation}
holds, then the theory is gauge invariant in second order. Note that the
distribution $T_2=R_2 + N_2 - R'_2$ then fulfils (\ref{gau}), too.
The local terms on the right-hand side of (\ref{gau}), which come
from the causal splitting, are called anomalies. The ordinary axial
anomalies in the standard model are of the same kind, they appear in the
third order triangle diagrams with axial vector couplings to fermions [10].

We consider the following example:
In the commutator $[T_{1/1}^\mu(x),T_1(y)]$ appears the term
\begin{equation}
-g^2 m :u(x)\Phi(x)[\partial_\mu \phi(x),\phi(y)]A_\nu(y)A^\nu(y):=
ig^2m:u(x)\Phi(x)A_\nu(y)A^\nu(y):\partial^\mu D_{m_H} (x-y).
\label{an}
\end{equation}
After splitting this causal distribution the Pauli-Jordan distribution
$D_{m_H}$ is replaced by the retarded distribution $D^{ret}_{m_H}$.
If we calculate now the divergence of (\ref{an}), we get an anomaly
\begin{equation}
\frac{A_1}{2}= ig^2m:u\Phi A_\nu A^\nu :\delta (x-y), \label{exan}
\end{equation}
because
\begin{equation}
\partial_\mu^x \partial_x^\mu D^{ret}_m (x-y) = - m^2 D^{ret}_m (x-y) +
\delta(x-y).
\end{equation}
The terms with $x$ and $y$ interchanged lead to the same contribution.
But in the causal distribution $D_2=[T_1(x),T_1(y)]$ appears the term
\begin{equation}
-g^2:A_\mu(x)\Phi(x)[\partial^\mu \phi(x),\partial^\nu \phi(y)]A_\nu
(y)\Phi(y):
\end{equation}
\begin{equation}
=-ig^2:A_\mu(x)A_\nu(y) \Phi(x)\Phi(y):\partial^\mu_x
\partial^\nu_x D_{m_H}(x-y)
\end{equation}
which has singular degree $\omega=0$ [2,3] and therefore allows a
normalization term in the splitted distribution
\begin{equation}
\partial_x^\nu\partial^\mu_x D^{ret}(x-y) \rightarrow
\partial_x^\nu\partial^\mu_x D^{ret}(x-y)+Cg^{\mu \nu} \delta(x-y).
\end{equation}
Since
\begin{equation}
d_Q(:\Phi^2 A_\mu A^\mu:C\delta(x-y)) = 2iCm:u\Phi A_\mu A^\mu:\delta(x-y)+...
\end{equation}
we can compensate the anomaly (\ref{exan}) by choosing $C=ig^2$.
In this way we obtain the quadrilinear couplings of the theory as
normalization terms in higher orders.
We give here the complete list of all normalization terms for tree diagrams
in second order:
\begin{equation}
N_1=ig^2:A_\mu A^\mu \Phi^2:\delta(x-y)
\end{equation}
\begin{equation}
N_2=ig^2:A_\mu A^\mu \phi^2:\delta(x-y)
\end{equation}
\begin{equation}
N_3=-ig^2\frac{m_H^2}{4m^2}:\Phi^4:\delta(x-y)
\end{equation}
\begin{equation}
N_4=ig^2(\frac{m_H^2}{m^2}+3 \delta_4):\phi^2\Phi^2:\delta(x-y)
\end{equation}
\begin{equation}
N_5=ig^2 \lambda' :\phi^4:\delta(x-y) \quad , \quad \lambda' \quad {\mbox{still
free}}
\end{equation}
The remaining free parameters $\delta_4$ and $\lambda'$ can be determined
by considering the anomalies $\sim \delta(x-z)\delta(y-z)$
of tree diagrams in third order.
They arise in the splitting of terms in
\begin{equation}
D_{3/1}^\mu(x,y,z)=[T_{1/1}^\mu(x),T_2(y,z)]+...
\end{equation}
where $T_{1/1}^\mu$ (\ref{te11})
gets contracted with a normalization term $N_{1-5}$ in
$T_2$.
Considering all anomalies $\sim :u\Phi\phi^3:$, gauge invariance requires
\begin{equation}
2\lambda'=\frac{m_H^2}{m^2}+3 \delta_4,
\end{equation}
and from the anomalies $\sim :u\phi\Phi^3:$ we obtain
\begin{equation}
\delta_4=-\frac{m_H^2}{2m^2},
\end{equation}
in agreement with (\ref{agree}).

Besides some basic assumptions concerning simplicity (42),
we have constructed the theory with the help of a guiding principle,
namely perturbative quantum gauge invariance, which, after construction,
is manifest in our approach.

\newpage
\vskip 1cm
{\it References}\vskip 1cm

[1] Higgs P W 1964 {{\em Phys.Rev.Lett. {\bf {12}}} 132

[2] Epstein H, Glaser V 1973 {\em{Ann.Inst.Poincar\'e A {\bf{29}}}} 211

[3] Scharf G 1995 {\em{Finite quantum electrodynamics, the causal approach}}

\hskip 0.5cm (second edition, Berlin, Heidelberg, New York: Springer)

[4] Malgrange B 1960 {\em{Seminaire Schwartz \bf{21}}}

[5] Krahe F 1996 {\em{Acta Physica Polonica B}}, to appear

[6] D\"utsch M, Hurth T , Scharf G 1995 {\em{Nuov. Cim. \bf{108A}}} 737

[7] Razumov A V, Rybkin G N 1990 {\em{Nucl.Phys. B \bf{332}}} 209

[8] Bognar J 1974 {\em{Indefinite inner product spaces}} (Berlin: Springer)

[9] D\"utsch M, Scharf G, hep-th/9612091

[10] Aste A, D\"utsch M, Scharf G, hep-th/9702053

[11] Cheng T P, Li L F 1984 {\em{Gauge theory of elementary particle physics}}

\hskip 0.5 cm (Oxford University Press)

\end{document}